\documentclass{article}  
\usepackage{abstract_2e} 

\usepackage{times}
\usepackage{epsfig}

\begin{document}  


\title{Retrieval of Planetary Rotation and Albedo from DSCOVR data}


\author{S.R. Kane}
\affil{Department of Physics \& Astronomy, San Francisco State
  University, 1600 Holloway Avenue, San Francisco, CA 94132, USA
  (skane@sfsu.edu)}
\author{S.D. Domagal-Goldman, J.R. Herman}
\affil{NASA Goddard Space Flight Center, Greenbelt, MD 20771, USA}
\author{T.D. Robinson}
\affil{Department of Astronomy and Astrophysics, University of
  California, Santa Cruz, CA 95064, USA}
\author{A.R. Stine}
\affil{Department of Earth \& Climate Sciences, San Francisco State
  University, 1600 Holloway Avenue, San Francisco, CA 94132, USA}


\runningtitle{Retrieval of Planetary Rotation and Albedo from DSCOVR
  data}

\titlemake  

\begin{abstracttext}

\section*{Introduction}

The field of exoplanets has rapidly expanded from the exclusivity of
exoplanet detection to include exoplanet characterization. This has
been enabled by developments such as the detection of
terrestrial-sized planets and the use of transit spectroscopy to study
exoplanet atmospheres. The studies of rocky planets presently being
undertaken are leading the path towards the direct imaging of
exoplanets and the development of techniques to extract their
intrinsic properties. The importance of properties such as rotation,
albedo, and obliquity is significant since they are key input
parameters for Global Climate Models used to determine surface
conditions, such as the models of Forget \& Lebonnois (2013). Thus, a
complete characterization of exoplanets for understanding exoclimates
requires the ability to obtain measurements of these key planetary
parameters.

The retrieval of planetary albedos and rotation rates from highly
undersampled imaging data can be informed by the use of climate
satellites designed to study the Earth's atmosphere. The Deep Space
Climate Observatory (DSCOVR) provides a unique opportunity to test
such retrieval methods using data for the sunlit surface of the
Earth. The high-resolution images can be deconvolved to match a
variety of expected exoplanet mission requirements and the relatively
high-cadence can be modified to design an effective observing
strategy. Our modeling of the DSCOVR data will provide an effective
baseline from which to develop tools that can be applied to a variety
of exoplanet imaging data.

\section*{The DSCOVR Mission}

The DSCOVR mission\footnote{http://www.nesdis.noaa.gov/DSCOVR/} is
primarily operated by the National Oceanic and Atmospheric
Administration (NOAA). DSCOVR was successfully launched on February
11, 2015 via a SpaceX Falcon 9 rocket. The satellite reached the L1
Lagrange point between the Earth and the Sun on June 8, 2015. The
satellite orbits the L1 point with a period of $\sim$6 months
resulting in an Earth-Sun viewing angle that varies in the range 4--15
degrees.

\begin{figure}[htb]
  \psfig{file=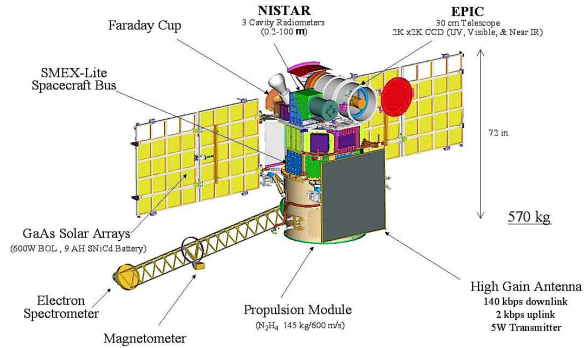,width=7.8cm}
  \caption{\em Earth-facing side of the DSCOVR satellite, showing the
    configuration of the instrumentation including the NISTAR and EPIC
    instruments described in this project.}
  \label{dscovr}
\end{figure}

\begin{figure*}[htb]
  \begin{tabular}{cc}
    \psfig{file=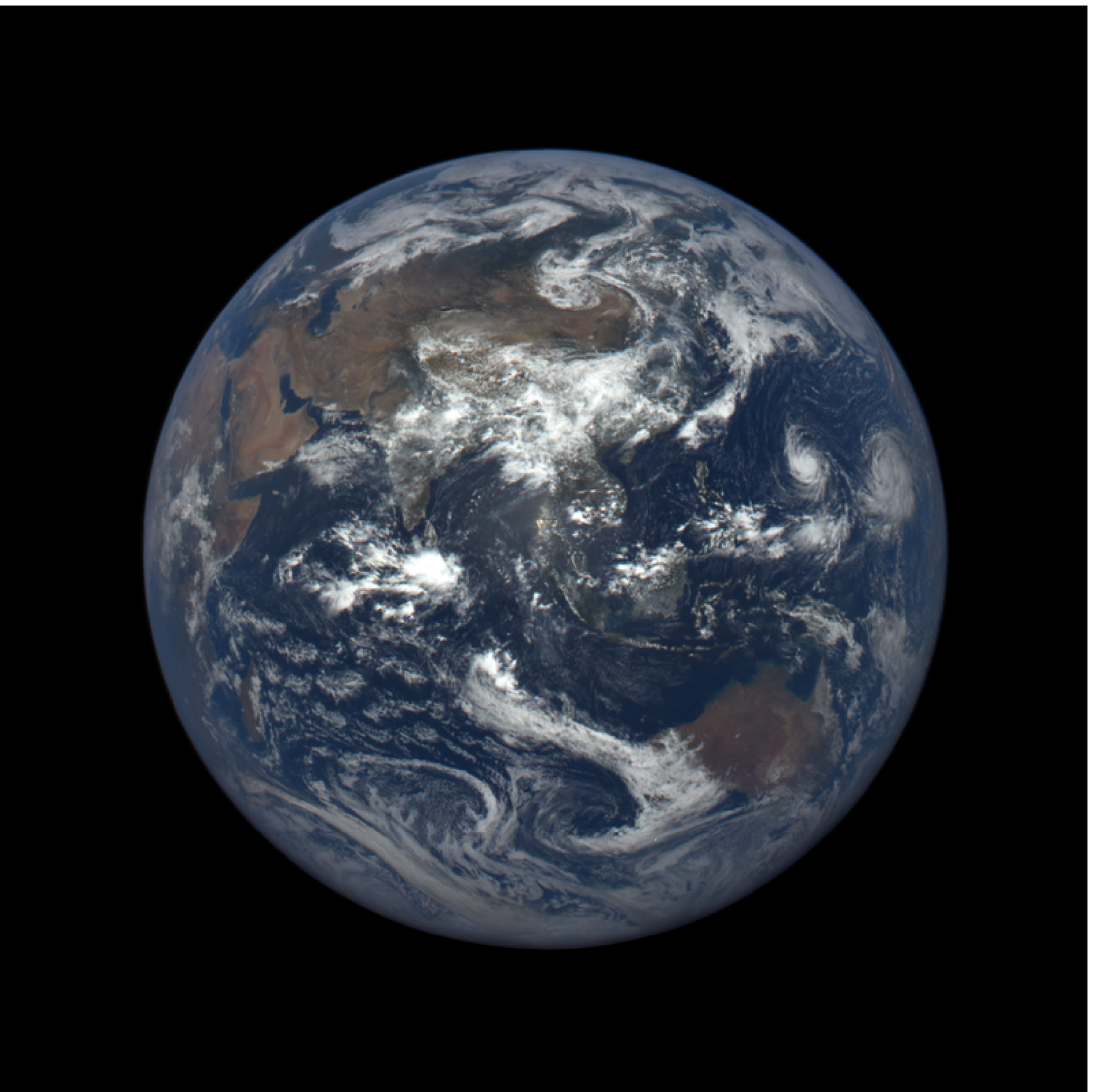,width=7.8cm} &
    \psfig{file=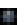,width=7.8cm}
  \end{tabular}
  \caption{\em Left: EPIC image of the Earth viewed from the L1
    Lagrange point. The image was acquired on August 17, 2015 at
    05:42:50~UT. Right: The same image as that shown at left after it
    has been deconvolved to 5$\times$5 pixel resolution.}
  \label{epic}
\end{figure*}

The details of DSCOVR hardware are shown in Figure~\ref{dscovr}. There
are two NASA instruments on board DSCOVR that are used to monitor the
Earth in detail: the National Institute of Standards and Technology
Advanced Radiometer (NISTAR) and the Earth Polychromatic Imaging
Camera (EPIC). NISTAR consists of three active cavity radiometers and
one silicon photodiode. The instrument uses three wide spectral bands
to study the reflectance of the Earth: total radiation in the range
0.2--100 microns, reflected solar radiation in the range 0.2--4
microns, and reflected near-infrared radiation in the range 0.7--4
microns. The combination of these data enables monitoring of changes
in Earth's total radiation budget.

EPIC is a 30~cm aperture f/9.6 Cassegrain telescope whose purpose is
to take images of the sunlit side of the Earth. The EPIC field-of-view
is 0.61$^\circ$ which, when coupled with the 2048$\times$2048 CCD,
provides an angular resolution of 1.07~arcsec, or $\sim$25~km/pixel
surface resolution at a latitude 60$^\circ$ from the equator. A sample
image from EPIC is shown in Figure~\ref{epic} (left). The EPIC data
are acquired from the spectroradiometer 10 narrowband channels in the
range 317.5--780~nm with an exposure time of 40~ms. These passbands
allow the investigation of specific surface and atmospheric features,
such as O$_3$, SO$_2$, aerosols at short wavelengths, and
clouds/vegetation at long wavelengths.

The NISTAR and EPIC data products will be made publicly available soon
after they are acquired. All color images from EPIC are provided in
high-resolution png format. The EPIC Level 1 (calibrated) individual
wavelength images will also be provided when processing is
complete. The data products may be acquired from the NASA Langley
Atmospheric Science Data Center\footnote{https://eosweb.lac.nasa.gov}.

\section*{The Exoplanet Connection}

The data from the DSCOVR mission provides a unique opportunity to
monitor the Earth as an exoplanet for an extended period of time. A
critical aspect of informing mission design for future NASA exoplanet
imaging instruments is determining the limits of planetary parameter
retrieval from severaly undersampled data. Studies using limited
datasets of Earth images have been performed by Cowan \& Strait (2013)
and Pall\'e et al. (2008). Our team are addressing this issue by
deconvolving EPIC images to resolutions of only several pixels, with
the inclusion of gaussian filters to account for dispersion
effects. Figure~\ref{epic} (right) shows an example of a 5$\times$5
pixel deconvolved image produced from a full-resolution EPIC image. We
generate time series for each pixel in the deconvolved images and use
fourier analysis techniques to extract periodic behavior due to
planetary rotation, weather patterns, and surface terrain. We further
measure the albedo from each deconvolved image which is then compared
with the integrated radiance measurements from NISTAR
observations. The time variance for each of the pixels, in particular
the asymmetry of the data from each hemisphere, are also used to place
constraints on the obliquity of the planet. Through this methodology,
we are creating a grid of retrieval rates as a function of image
resolution, observing cadence, passband, etc.

There are various caveats that we are also taking care to account for
in our analysis. Perhaps the most obvious is that the data we are
using is specific to the Terran atmosphere. Results from the Kepler
mission are revealing a continuum of terrestrial planetary sizes and
masses (Burke et al. 2015), and there is likely a similar continuum of
atmospheric parameters, including composition, mass, and dynamics. We
thus are carefully considering how the results of this work will vary
with different planetary and atmospheric properties. Due to the
orbital motion of DSCOVR around L1, our analysis takes into account
both the change in the Earth viewing angle and the angular size of the
Earth-disk in the field-of-view. Finally, the orbital inclination of
an exoplanet relative to line of sight will play a significant role in
the observed analogous signature. The planetary phase of the exoplanet
as it orbits the host star will also modify the signature. Both of
these will be included in our simulations by overlaying the night-side
of the planet onto the Earth-disk prior to deconvolution.

\section*{Conclusions}

Exoplanet characterization is currently dominated by measurables such
as orbital elements and the planetary mass and radius. The detection
of basic atmospheric constituents has come into reach for those
transiting planets sufficiently close to their host star with a
relatively large scale height. Future direct imaging missions will
open further doors into extracting the fundamental instrinsic
properties of the planet that are key components in the modeling of
their climates. These include the planetary rotation and obliquity,
and atmospheric albedo. The parameter extraction techniques we are
developing will be used to inform mission design constraints on the
required cadence and image resolution needed to constrain these
important planetary parameters.

\section*{References}

\noindent -- Burke, C.J., et al. 2015, ApJ, 809, 8

\noindent -- Cowan, N.B., Strait, T.E. 2013, ApJ, 765, L17

\noindent -- Forget, F., Lebonnois, S. 2013, Comparative Climatology
of Terrestrial Planets, Stephen J. Mackwell, Amy A. Simon-Miller,
Jerald W. Harder, and Mark A. Bullock (eds.), University of Arizona
Press, Tucson, 610 pp., p.213-229

\noindent -- Pall\'e, E., Ford, E.B., Seager, S.,
Monta\~n\'es-Rodr\'iguez, P., Vazquez, M. 2008, ApJ, 676, 1319

\end{abstracttext}

\end{document}